\documentclass{ws-procs9x6}            
\begin{document}
\title{Positronium on the light front}

\author{Kaiyu Fu,$^*$ Hengfei Zhao,$^\dagger$ and Xingbo Zhao$^\$$}

\address{Institute of Modern Physics, Chinese Academy of Sciences, Lanzhou 730000, China\\
and School of Nuclear Science and Technology, University of Chinese Academy of Sciences,
Beijing 100049, China\\
$^*$E-mail: kaiyufu@impcas.ac.cn\\
$^\dagger$E-mail: zhaohengfei@impcas.ac.cn\\
$^\$$E-mail: xbzhao@impcas.ac.cn}

\author{James P. Vary$^\ddagger$}

\address{Department of Physics and Astronomy, Iowa State University,\\ Ames, IA 50011, USA\\
$^\ddagger$E-mail: jvary@iastate.edu\\
$($BLFQ Collaboration$)$}

\begin{abstract}
Basis Light-front Quantization (BLFQ) is a newly developed nonperturbative approach, aiming at solving relativistic bound systems based on the Hamiltonian formalism of light-front dynamics. In this work, we introduce its application to the positronium system at strong coupling, $\alpha=0.3$, with a dynamical photon mediating the interaction between the positron and the electron. Nonperturbative mass renormalization is needed to cancel the fermion self-energy divergence. Here, we present the resulting mass spectrum, light-front wave functions (LFWFs), and the photon distribution inside positronium.
\end{abstract}

\keywords{BLFQ; positronium; light-front wave functions; photon distribution.}

\bodymatter

\section{Introduction}

Nonperturbative calculations are needed for understanding hadron strunctures in terms of fundamental quark and gluon degrees of freedom. BLFQ \cite{PhysRevC.81.035205} is developed as a nonperturbative approach to bound states in a quantum field theory such as QCD. It combines the advantages of light-front dynamics\cite{BRODSKY1998299} and modern developments of $ab\; initio$ nuclear structure calculations\cite{PhysRevLett.84.5728,PhysRevC.62.054311}. BLFQ solves the Hamiltonian eigenvalue equation of the bound state in light-front coordinates with light-front gauge $A^+=0$. The resulting mass spectra and light-front wave functions (LFWFs) encode all the informations of the system.  
BLFQ has been successfully applied to the anomalous magnetic moment of the electron\cite{Honkanen,Zhao}, positronium\cite{PhysRevD.91.105009}, quarkonium\cite{y.li,lan}, meson\cite{s.jia,lan1,lan2}, and proton\cite{s.xu} systems. In this work, we apply BLFQ to the positronium system in QED while retaining a photon as a dynamical degree of freedom. Positronium is an ideal test case for the inclusion of the higher Fock sector since QED is simpler than QCD. Fock sector dependent renormalization\cite{PhysRevD.77.085028,PhysRevD.86.085006,ZHAO201465} is needed to cancel the fermion self-energy divergence. Here we present the resulting LFWF, mass spectrum, and photon distribution function.


\section{Basis Light-Front Quantization}

In BLFQ, we solve the eigenvalue equation $(P^{-}P^+ -P^{\perp 2})|\psi\rangle= M^{2}|\psi\rangle$, where $(P^-, P^+, P^\perp)$ is the energy momentum four-vector operator and $M$ is the invariant mass, $\psi$ is the mass eigenstate. In light-front gauge, $P^2 = P^{-}P^+ -P^{\perp 2}$ is the light-front Hamiltonian with the eigenvalue of mass square. $P^+$ and $P^\perp$ are the total longitudinal and transverse momenta of the system.
\begin{equation}
\begin{aligned}
P^-= &\int \mathrm{d}^{2} x^{\perp} \mathrm{d} x^{-} \frac{1}{2} \bar{\Psi} \gamma^{+} \frac{m_{e}^{2}+\left(i \partial^{\perp}\right)^{2}}{i \partial^{+}} \Psi+\frac{1}{2} A^{j}\left(i \partial^{\perp}\right)^{2} A^{j} \\ &+e j^{\mu} A_{\mu}+\frac{e^{2}}{2} j^{+} \frac{1}{\left(i \partial^{+}\right)^{2}} j^{+} ,
\end{aligned}
\end{equation}
which is derived from the QED Lagrangian in light-front coordinates via the Legendre transformation. 
In BLFQ, the Hamiltonian eigenvalue equation is expressed in a truncated discretized basis and solved numerically by diagonalizing the resulting Hamiltonian matrix. 
We first expand the positronium state in a truncated Fock space,
$|e^+ e^-\rangle_{phys}= a|e^+ e^-\rangle+ b|e^+ e^- \gamma\rangle$, 
where $|e^+ e^-\rangle$ and $|e^+ e^-\gamma\rangle$ are Fock sectors that we elect to retain. We then expand these Fock sectors into the three dimensional discretized basis. The transverse direction of the basis is a 2-dimensional harmonic oscillator, $\phi^{b}_{n,m}(\vec{p}_\perp)$, where $n$ is the radial quantum number, $m$ is the angular quantum number and b is a scale parameter. The longitudinal direction is the plane wave which is discretized by putting it in a ``box'' of Length $L$ and applying periodic boundary conditions for bosons and antiperiodic boundary conditions for fermions. So, the longitudinal momentum $p^+=\frac{2\pi}{L}j$, where j is an integer for bosons or a half-integer for fermions.  
The basis is made finite by restricting the quantum number in each basis state according to
 $\sum_{i}\left(2 n_{i}+\left|m_{i}\right|+1\right)  \leq N_{max } $ and $ \sum_{i} j_{i} =K, $
where the sum is over Fock particles in each Fock basis. So, the longitudinal momentum fraction is $x=\frac{p^+}{P^+}=\frac{j}{K} $. $N_{max}$ and $K$ are basis truncation parameters. Large $N_{max}$ and $K$ mean high ultraviolet cutoff and low infrared cutoff.

\section{Numerical results}

\begin{figure}
\begin{center}
\includegraphics[width=4in]{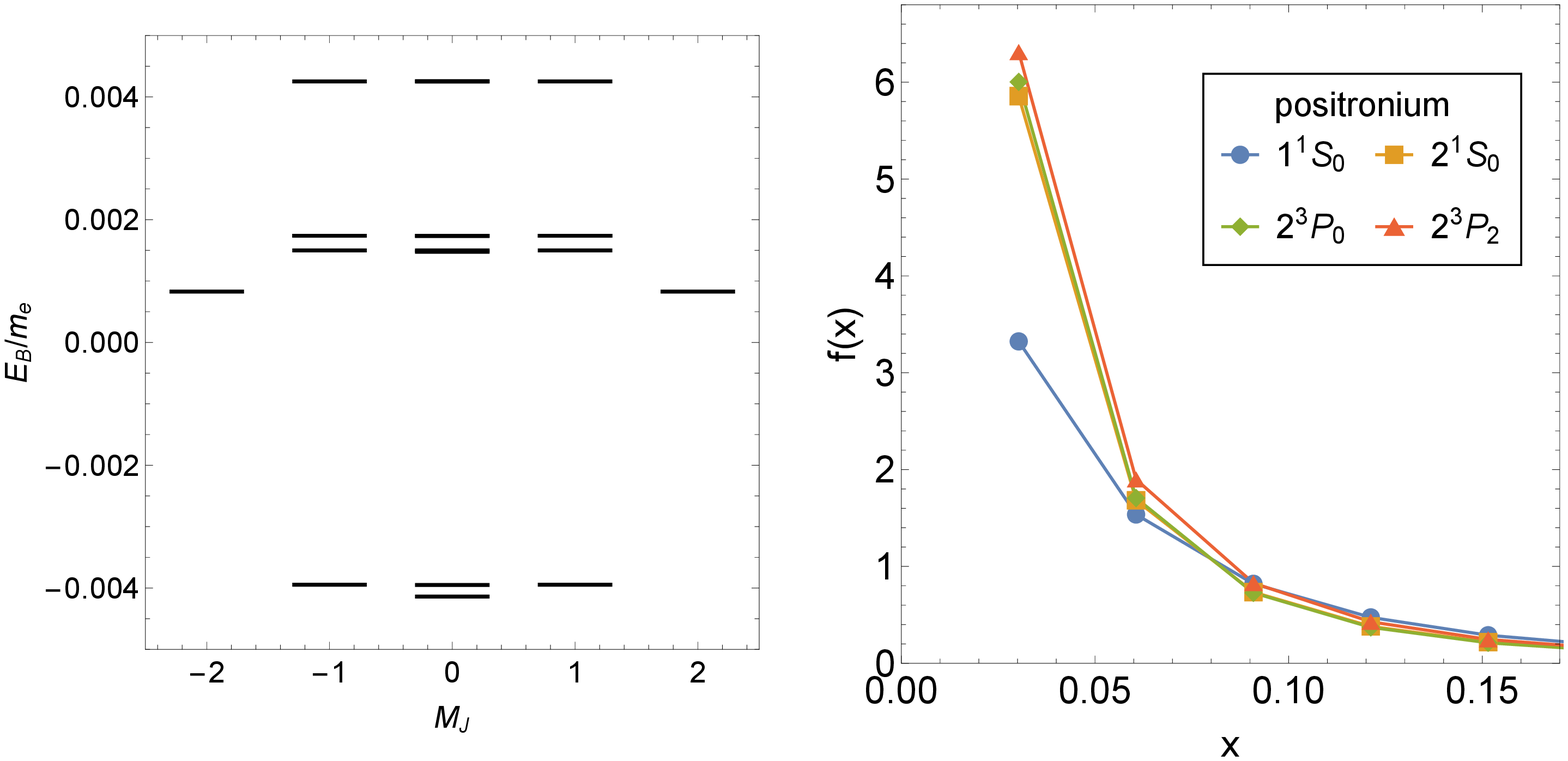}
\end{center}
\caption{Photon distribution function and interaction energy spectrum of positronium at $N_{max}=K-1=32$, $\alpha=0.3$. $M_J$ is the total angular momentum projection of the system and $m_e$ is the physical electron mass. }
\label{aba:fig1}
\end{figure}

\begin{figure}
\begin{center}
\includegraphics[width=1.6in]{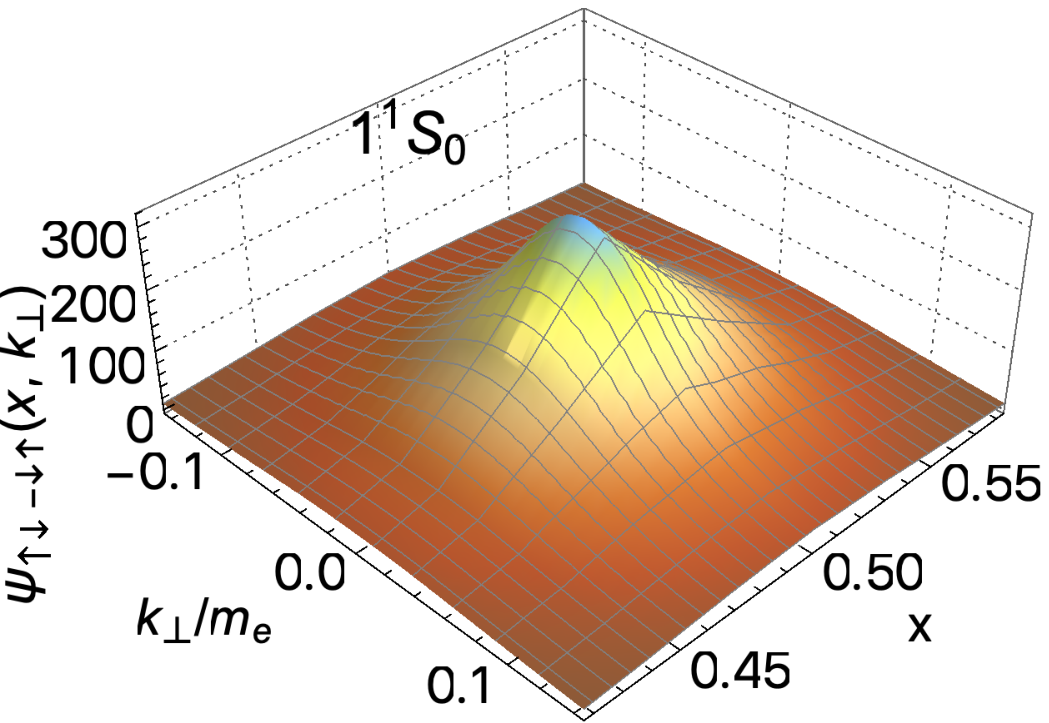} 
\includegraphics[width=1.6in]{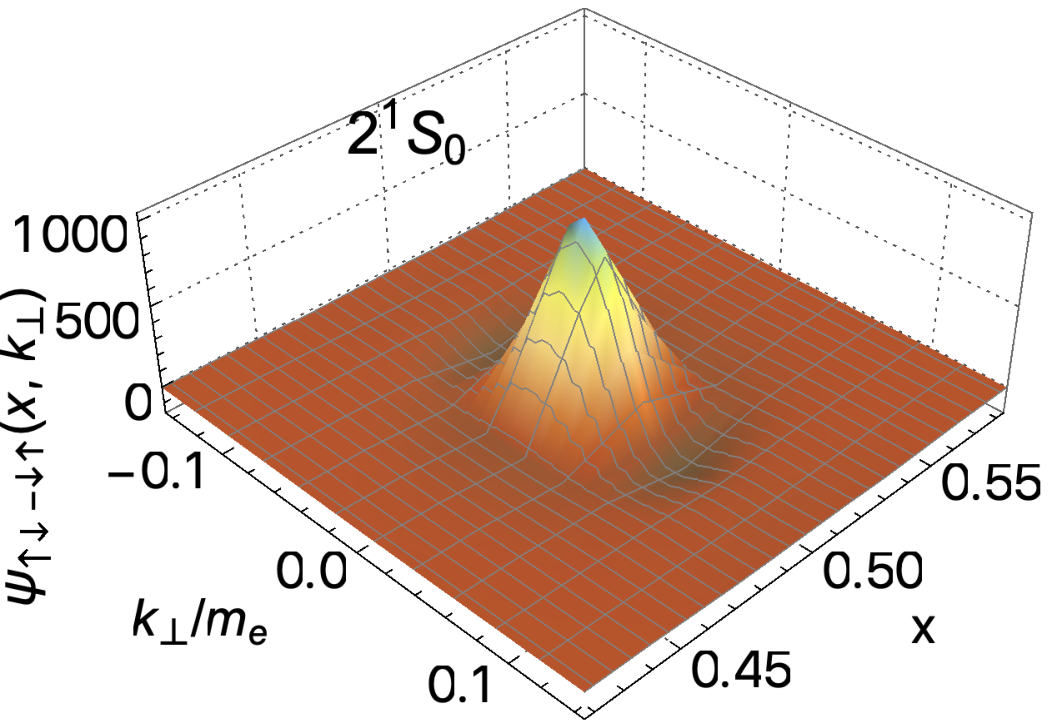}
\includegraphics[width=1.6in]{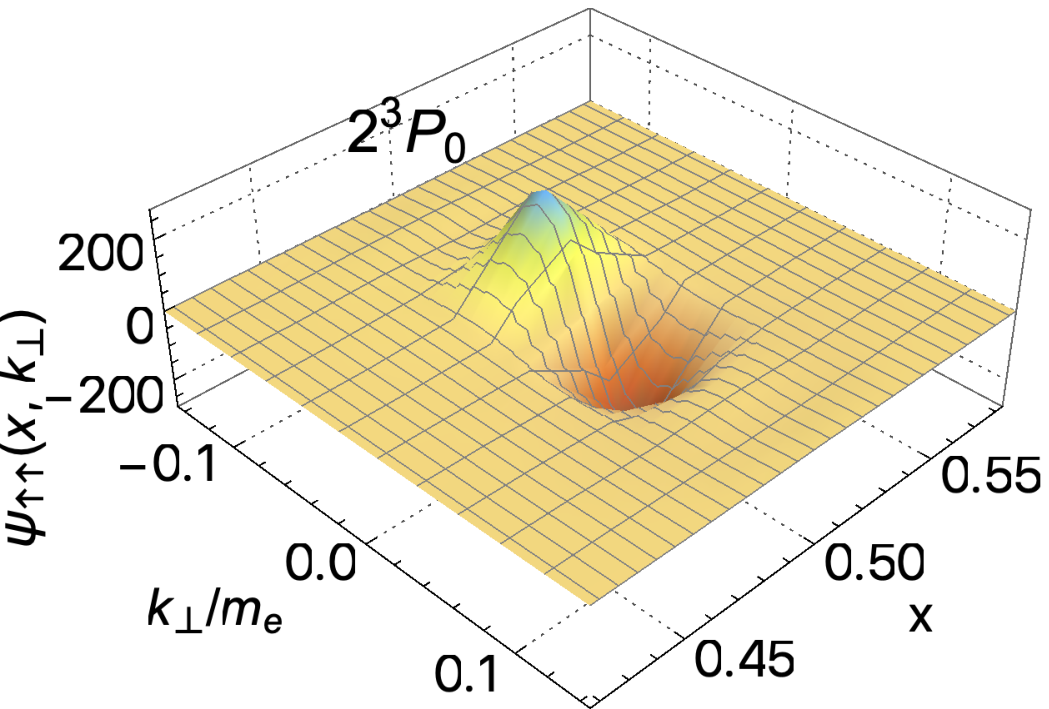}
\includegraphics[width=1.6in]{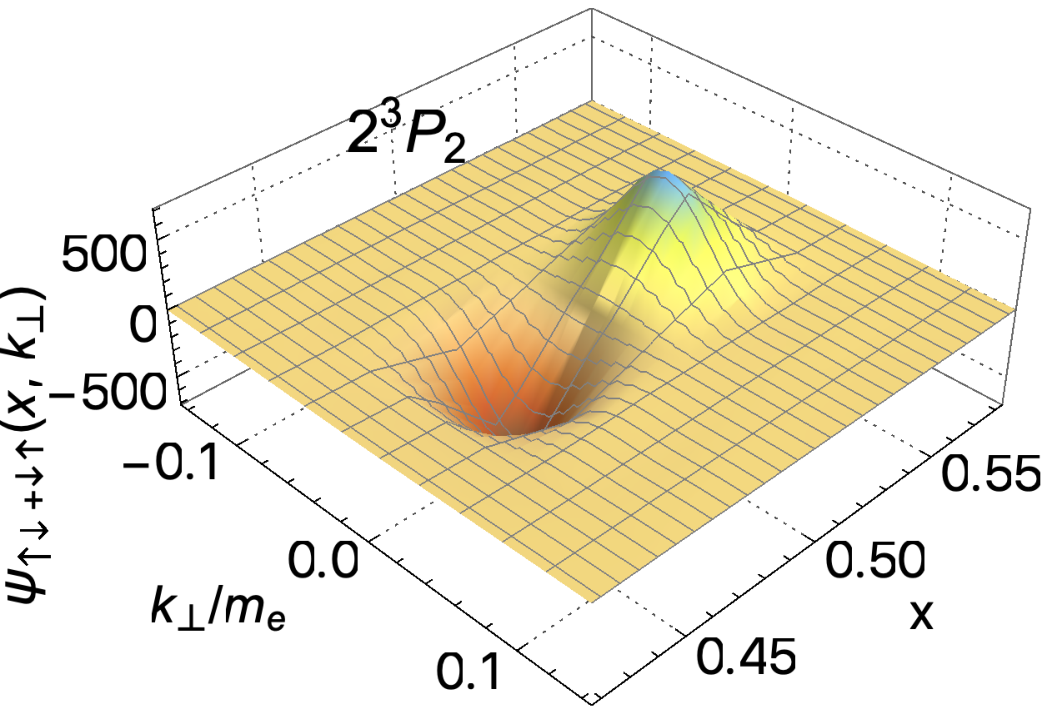}
\end{center}
\caption{LFWFs of $|e^+ e^-\rangle$ at $N_{max}=K-1=32$, $\alpha=0.3$. Here $x$ is the relative longitudinal momentum fraction and $k^\perp$ is the relative transverse momentum.}
\label{aba:fig2}
\end{figure}

In Fig.~\ref{aba:fig1}, $E_B$ is the interaction energy defined as the difference between the invariant mass of a bound state and the sum of masses of its component particles.
The left panel in Fig.~\ref{aba:fig1} is the energy spectrum with the identified lowest eight states. Using the approximate degeneracy and relations between parity and quantum numbers\cite{,y.li}, we can identify the lowest eight states. In the $M_J=0$ column, the ordering of these states, from down to up, is $1^1S_0$, $1^3S_1$, $2^1S_0$, $2^3S_1$, $2^3P_0$, $2^3P_1$, $2^1P_1$, $2^3P_2$. Fig.~\ref{aba:fig2} displays corresponding $|e^+ e^-\rangle$ LFWFs of $M_J =0$. The spectrum and LFWFs of $|e^+ e^-\rangle$ reasonably agree with those from the effective one-photon-exchange calculation\cite{PhysRevD.91.105009}. In addition, we obtain the LFWF's $|e^+ e^-\gamma\rangle$ components which is not included in Ref.\cite{PhysRevD.91.105009} and gives direct access to the photon content in positronium.
$f(x)$ is the photon distribution function of positronium,
\begin{equation}
f(x)=\int |\psi(\beta,x)|^2 d\beta,
\end{equation}
where $x$ is the longitudinal momentum fraction of the photon and $\beta$ represents the remaining variables in the LFWF of $|e^+ e^-\gamma\rangle$. The right panel in Fig.\ref{aba:fig1} shows that excited states have more photon content than the ground state in the small $x$ region.

\section{Conclusions }

We apply BLFQ to the positronium system using the light-front QED Hamiltonian while retaining a photon as a dynamical degree of freedom. Sector-dependent renormalization appears to successful for canceling divergences. We obtain the reasonable spectrum and LFWFs along with the information of the photon content in positronium at strong coupling. By adding an effective confining potential between fermions, this positronium calculation can be straightforwardly extended to heavy quarkonium and light meson systems.

\section*{Acknowledgments}

We thank J. Lan, S. Xu, Y. Li, and C. Mondal for many insightful discussions. This work of X. Z. is supported by new faculty startup funding by the Institute of Modern Physics, Chinese Academy of Sciences under the Grant No.~Y632030YRC. J. P. V. is supported by the Department of Energy under Grants No.~DE-FG02-87ER40371 and No.~DE-SC0018223 (SciDAC4/NUCLEI). A portion of the computational resources were provided by the National Energy Research Scientific Computing Center (NERSC), which is supported by the Office of Science of the U.S. Department of Energy under Contract No. DE- AC02-05CH11231.

\bibliographystyle{ws-procs9x6} 


\end{document}